\documentclass[conference]{IEEEtran}

\usepackage{cite}
\usepackage{graphicx}
\usepackage{psfrag}
\usepackage{url}
\usepackage{amsmath}
\usepackage{array}
\usepackage{amssymb}
\usepackage{amsthm}
\usepackage{mathtools}
\usepackage{amsfonts}
\usepackage{graphicx}
\usepackage{epstopdf}
\usepackage{algorithm}
\usepackage{algorithmic}

\usepackage{setspace}
\usepackage{amscd}
\usepackage{mathrsfs}
\usepackage{epsfig}
\usepackage{color}
\usepackage{textcomp}
\usepackage{comment}
\usepackage{xcolor} 
\usepackage{pgf}	
\usepackage{pgfplots}
\usepackage{tikz}
\usetikzlibrary{decorations.pathreplacing}
\usetikzlibrary{decorations.markings}
\usetikzlibrary{patterns}
\usepackage[outline]{contour}
\contourlength{1.2pt}
\pgfplotsset{compat=1.9} 
\usetikzlibrary{plotmarks}
\usetikzlibrary{spy}
\usetikzlibrary{calc}
\usetikzlibrary{shapes,plotmarks,positioning,fit,chains,tikzmark,arrows,hobby,arrows.meta}
\usetikzlibrary{decorations.pathreplacing,calligraphy}

\usepackage[utf8]{inputenc}
\usepackage[T1]{fontenc}

\IEEEoverridecommandlockouts

\usepackage[toc, notranslate, acronym]{glossaries}
\usepackage{glossaries-extra}
\setabbreviationstyle[acronym]{long-short}
\newacronym{4g}{4G}{fourth generation}
\newacronym{5g}{5G}{fifth generation}
\newacronym{6g}{6G}{sixth generation}
\newacronym{fwa}{FWA}{fixed wireless access}
\newacronym{afr}{AFR}{amplify-and-forward repeater}
\newacronym{mimo}{MIMO}{multiple-input multiple-output}
\newacronym{dmimo}{D-MIMO}{distributed multiple-input multiple-output}
\newacronym{siso}{SISO}{single-input single-output}
\newacronym{simo}{SIMO}{single-input multiple-output}
\newacronym{miso}{MISO}{multiple-input single-output}
\newacronym{iid}{i.i.d.}{independent and identically distributed}
\newacronym{bs}{BS}{base station}
\newacronym{ue}{UE}{user equipment}
\newacronym{ap}{AP}{access point}
\newacronym{upa}{UPA}{uniform planar array}
\newacronym{los}{LoS}{line-of-sight}
\newacronym{awgn}{AWGN}{additive white Gaussian noise}
\newacronym{isp}{ISP}{Internet service provider}
\newacronym{mmse}{MMSE}{minimum-mean-square error}
\newacronym{ls}{LS}{least-squares}
\newacronym{mse}{MSE}{mean-square error}
\newacronym{rmse}{RMSE}{root mean square error}
\newacronym{zf}{ZF}{zero-forcing}
\newacronym{mrc}{MRC}{maximum-ratio combiner}
\newacronym{mrt}{MRT}{maximum-ratio transmission}
\newacronym{se}{SE}{spectral efficiency}
\newacronym{snr}{SNR}{signal-to-noise ratio}
\newacronym{sinr}{SINR}{signal-to-interference-plus-noise ratio}
\newacronym{rf}{RF}{radio frequency}
\newacronym{ofdm}{OFDM}{orthogonal frequency division multiplexing}
\newacronym{fcch}{FCCH}{frequency correction channel}
\newacronym{fb}{FB}{frequency correction burst}
\newacronym{gsm}{GSM}{global system for mobile}
\newacronym{crlb}{CRLB}{Cram{\'e}r-Rao lower bound}
\newacronym{fim}{FIM}{Fisher information matrix}
\newacronym{dft}{DFT}{discrete Fourier transform}
\newacronym{svd}{SVD}{singular value decomposition}
\newacronym{evd}{EVD}{eigenvalue decomposition}
\newacronym{cp}{CP}{cyclic prefix}
\newacronym{lo}{LO}{local oscillator}
\newacronym{tdd}{TDD}{time division duplexing}
\newacronym{ml}{ML}{maximum-likelihood}
\newacronym{csi}{CSI}{channel state information}
\newacronym{pcsi}{PCSI}{perfect channel state information}
\newacronym{cpu}{CPU}{central processing unit}
\newacronym{fgb}{FGB}{fixed grid of beams}
\newacronym{nls}{NLS}{non-linear least squares}
\newacronym{vco}{VCO}{voltage controlled oscillator}
\newacronym{pll}{PLL}{phase locked loop}
\newacronym{fdma}{FDMA}{frequency division multiple access}
\newacronym{urllc}{URLLC}{ultra-reliable low latency communication}
\newacronym{embb}{eMBB}{enhanced mobile broadband}
\newacronym{mmtc}{mMTC}{massive machine type communications}
\newacronym{mtc}{MTC}{machine type communication}
\newacronym{iot}{IoT}{Internet-of-Things}
\newacronym{cs}{CS}{compressed sensing}
\newacronym{amp}{AMP}{approximate message parsing}
\newacronym{roc}{ROC}{receiver operating characteristic}
\newacronym{3gpp}{3GPP}{3rd Generation Partnership Project}
\newacronym{vr}{VR}{virtual reality}
\newacronym{ar}{AR}{augumented reality}
\newacronym{un}{UN}{United Nations}
\newacronym{rru}{RRU}{remote radio unit}
\newacronym{mr}{MR}{maximum-ratio}
\newacronym{fc}{FC}{fully coherent}
\newacronym{fd}{FD}{full-duplex}
\newacronym{fnc}{FNC}{fully non-coherent}
\newacronym{pcnc}{PC}{partially coherent}
\newacronym{wmmse}{WMMSE}{weighted minimum-mean-square-error}
\newacronym{bcd}{BCD}{block coordinate descent}
\newacronym{qcqp}{QCQP}{quadratically constrained quadratic program}
\newacronym{sic}{SIC}{successive interference cancellation}
\newacronym{cinr}{CINR}{channel-to-noise-plus-interference-channel ratio}
\newacronym{hd}{HD}{high-definition}
\newacronym{lte}{LTE}{long-term evolution}
\newacronym{rach}{RACH}{random access channel}
\newacronym{prach}{PRACH}{physical random access channel}
\newacronym{rrc}{RRC}{radio resource control}
\newacronym{sucr}{SUCR}{strongest user collision resolution}
\newacronym{noma}{NOMA}{non-orthogonal multiple access}
\newacronym{psd}{PSD}{positive semi-definite}
\newacronym{cran}{C-RAN}{cloud radio access network}
\newacronym{comp}{CoMP}{Coordinated MultiPoint}
\newacronym{fdd}{FDD}{frequency division duplexing}
\newacronym{dsp}{DSP}{digital signal processor}
\newacronym{trp}{TRP}{transmission reception point}
\newacronym{leo}{LEO}{low-earth orbit}
\newacronym{cjt}{CJT}{coherent joint transmission}
\newacronym{ris}{RIS}{reflective intelligent surface}
\newacronym{ota}{OTA}{over-the-air}

\newcommand{\Brac}[1]{\left(#1\right)}

\newcommand{\Abs}[1]{\left\vert #1 \right\vert}
\newcommand{\Norm}[1]{\Vert #1 \Vert}

\newcommand{\rhoR}{\rho_{\text{R}}}
\newcommand{\rhoA}{\rho_{\text{A}}}
\newcommand{\rhoB}{\rho_{\text{B}}}

\newcommand{\tA}{{\text{A}}}
\newcommand{\tB}{{\text{B}}}
\newcommand{\tR}{{\text{R}}}

\newcommand{\tH}{\mathrm{H}}
\newcommand{\tT}{\mathrm{T}}

\newcommand{\CC}{\mathbb{C}}

\newcommand{\D}{\mathbf{D}}

\newcommand{\EE}{\mathbb{E}}

\newcommand{\f}{\mathbf{f}}

\newcommand{\g}{\mathbf{g}}

\newcommand{\h}{\mathbf{h}}
\let\H\undefined 
\newcommand{\H}{\mathbf{H}}

\newcommand{\rr}{\mathbf{r}}
\newcommand{\R}{\mathbf{R}}

\let\t\undefined
\newcommand{\t}{\mathbf{t}}

\newcommand{\w}{\mathbf{w}}
\newcommand{\W}{\mathbf{W}}

\newcommand{\x}{\mathbf{x}}

\newcommand{\y}{\mathbf{y}}
\newcommand{\Y}{\mathbf{Y}}

\newcommand{\CN}{\mathcal{CN}}

\tikzset{
	bs/.pic = {		
		\draw[line width = 1pt,-round cap] (0,0.4\R) -- (-0.2\R,-0.2\R);
		\draw[line width = 1pt,-round cap] (0,0.4\R) -- (0.2\R,-0.2\R);
		\draw[line width = 1pt] (-0.2\R,-0.2\R) -- (0.133\R,0);
		\draw[line width = 1pt] (0.2\R,-0.2\R) -- (-0.133\R,0);
		\draw[line width = 1pt,-round cap] (-0.133\R,0) -- (0.067\R,0.2\R);
		\draw[line width = 1pt,-round cap] (0.133\R,0) -- (-0.067\R,0.2\R);
		\draw[line width = 1pt,-round cap] (-0.213\R,0.4\R) -- (0.213\R,0.4\R);
		\draw[line width = 1pt,-round cap] \foreach \x in {-0.21, -0.14,...,0.22} {(\x\R,0.4\R) -- (\x\R,0.47\R)};
		\node (dim) at (0,0)  [align=center,minimum width=0.5\R,minimum height=1\R] {};
	},
}

\tikzset{
	user/.pic = {	
		\draw[rounded corners=0.02\R] (-0.09\R,-0.17\R) rectangle (0.09\R,0.17\R) ;	
		\draw[fill=gray] (-0.08\R,-0.12\R) rectangle (0.08\R,0.12\R);	
		\draw[rounded corners=0.005\R] (-0.04\R,0.14\R) rectangle (0.04\R,0.15\R);	
		\draw (0,-0.14\R) circle (0.012\R) ; 
		\node (dim) at (0,0)  [align=center,minimum width=0.2\R,minimum height=0.3\R] {};
	}
}

\title{Repeater-Aided Over-the-Air Phase Synchronization in Distributed MIMO}
\author{
\IEEEauthorblockN{Unnikrishnan Kunnath Ganesan$^*$, Sai Subramanyam Thoota$^{\dagger}$, 
Erik G. Larsson$^{\ddagger}$}
\IEEEauthorblockA{
$^{*}$Ericsson AB, Sweden (email: unnikrishnan.kunnath.ganesan@ericsson.com)\\
$^{\dagger}$Nokia, India (email: sai.subramanyam\_thoota@nokia.com) \\
$^{\ddagger}$Department of Electrical Engineering, Link\"oping University, 581 83 Link\"oping, Sweden (email: erik.g.larsson@liu.se)
}
\thanks{The work of E. G. Larsson was supported in part by ELLIIT, VR, and the KAW foundation.}
}

\begin{document}

\maketitle

\begin{abstract}
Phase synchronization of \glspl{ap} in a \gls{dmimo} system is critical to leverage the performance benefits of D-MIMO. 
Existing over-the-air phase synchronization methods assume that \glspl{ap} can communicate directly to perform necessary measurements. 
However, this assumption might not hold in scenarios where inter-\gls{ap} signaling is too weak for effective communication. 
To address this, in this paper, we propose a novel over-the-air calibration scheme  that uses repeater nodes to facilitate phase synchronization when direct \gls{ap} signaling is infeasible. 
We give the steps of the algorithm   for phase calibration in  closed form, and show how it enables \gls{cjt} by the \glspl{ap}. 
The framework expands the applicability of \gls{dmimo} systems to challenging environments, where existing over-the-air synchronization techniques fall short.

\end{abstract}


\glsresetall

\section{Introduction} \label{sec:introduction}
\subsection{Context and Motivation}
\Gls{dmimo} is a key technology for next-generation cellular networks and offers several advantages over traditional centralized MIMO systems~\cite{ngo2017cell,interdonato2019ubiquitous}. 
Multiple \glspl{ap} are distributed across the network, providing enhanced data rates, lower latency, and improved coverage, even in challenging environments.
Furthermore, \gls{dmimo} provides support for a wide range of users, from high-demand devices to low-power \gls{iot} devices, meeting the diverse needs of next-generation wireless systems~\cite{ngo2017cell}.

Any timing or frequency misalignment between \glspl{ap} causes relative phase offsets between them, which affects the coherent combining of the signals transmitted by them.
Hence,   time and frequency synchronization of distributed \glspl{ap} is critical to fully leverage the advantages of \gls{dmimo}. 
Moreover, the hardware of each \gls{ap} being independent, reciprocity calibration must also be addressed along with phase synchronization~\cite{nissel2022correctly,kaltenberger2010relative}. 
Wired synchronization solutions can align the \glspl{ap}; however, they lack scalability. 
In contrast, over-the-air phase synchronization, as explored in~\cite{rogalin2014scalable,ganesan2024beamsync,larsson2023phase,larsson2024massive}, facilitates scalable synchronization across multiple \glspl{ap} and achieves joint reciprocity-calibration of the APs, thereby enabling \gls{cjt}.

\begin{figure}
    \centering
    \begin{tikzpicture}[font=\scriptsize]
    \def\buildingWidth{3}
    \def\buildingDepth{2}
    \def\buildingHeight{4}
    
    \fill[gray!30] (0, 0) -- ++(\buildingWidth, 0) -- ++(0, \buildingHeight) -- ++(-\buildingWidth, 0) -- cycle;
    


    \draw[thick] (0, 0) -- ++(\buildingWidth, 0) -- ++(0, \buildingHeight) -- ++(-\buildingWidth, 0) -- cycle;
    
    \foreach \x in {0.5, 2} {
        \fill[blue!50] (\x, 2) rectangle ++(0.5, 0.5);
        \fill[blue!50] (\x, 3) rectangle ++(0.5, 0.5);
    }

    \fill[brown] (1.2, 0.05) rectangle ++(0.75, 1.5);

    
    \newdimen\R
    \R=3cm
    \draw (-1.5,1) pic (ap1) {bs} ;
    \draw (\buildingWidth+1.5,1) pic (ap2) {bs} ;

    \node  [align=center,above=0.01mm of ap1dim.south] {AP A \\ $(M_\text{A},\mathbf{t}_\text{A},\mathbf{r}_\text{A})$};
    \node  [align=center,above=0.01mm of ap2dim.south] {AP B \\ $(M_\text{B},\mathbf{t}_\text{B},\mathbf{r}_\text{B}$)};

    \filldraw[fill=red!30, draw=black, line width=1pt] (1,\buildingHeight) rectangle (2,\buildingHeight+0.5) node at (3,\buildingHeight+1) (AuxNode) {repeater R};
    
    \draw[thick] (1,\buildingHeight+0.5) -- (1,\buildingHeight+1);


    \fill (1,\buildingHeight+1.1) circle (0.05);

    \draw[thick] (1+0.1,\buildingHeight+1.1) arc[start angle=0, end angle=180, radius=0.1];

    \draw[thick] (1+0.2,\buildingHeight+1.1) arc[start angle=0, end angle=180, radius=0.2];

    \draw[thick] (1+0.3,\buildingHeight+1.1) arc[start angle=0, end angle=180, radius=0.3];

\end{tikzpicture}
    \caption{System model depicting weak signaling between two APs due to signal blockage by a tall building in between. }
    \label{fig:system_model}
\end{figure}
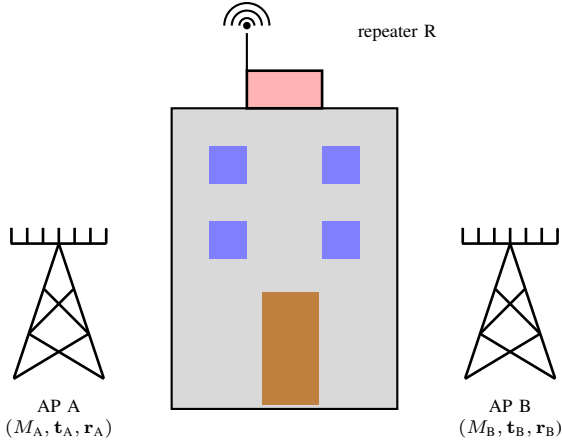

Over-the-air synchronization methods assume that \glspl{ap} can directly perform measurements on  one another. 
In this paper, we address a scenario where the inter-\gls{ap} signaling is too weak for direct measurements, even at higher transmission powers. 
This could occur, for instance, when severe shadow fading is caused by obstructions like tall buildings, illustrated in Fig.~\ref{fig:system_model}. 
However, each \gls{ap} may still communicate with an intermediary node in the form of a \textit{repeater} (see, e.g.,  \cite{sara2025repeater,ma2015channel,carvalho2025network} for a discussion of the role of repeaters in future systems). 
Repeaters have been traditionally employed to eliminate blind spots in massive MIMO systems, thereby enhancing data rates.
In this paper, we explore how repeaters can be utilized to improve the calibration of distributed \glspl{ap} in \gls{dmimo} systems.



\subsection{Contributions}
The specific contributions of our paper are: 
\begin{itemize}
    \item We propose a novel reciprocity calibration protocol for two \glspl{ap} when a repeater is placed to communicate between them.
    \item We derive a closed-form expression for the phase estimate, and illustrate the advantages of our proposed protocol using numerical simulations.
\end{itemize}
To the best of our knowledge, no prior literature has considered the use of an intermediate node (repeater)
to aid in over-the-air reciprocity calibration of two \glspl{ap} in \gls{dmimo}. 

\textbf{\textit{Notation:}} 
Bold lowercase and uppercase letters  denote vectors and matrices, respectively. 
The operations $(\cdot)^\mathrm{T}$ and  $(\cdot)^\mathrm{H}$ denote the transpose and Hermitian transpose, respectively. 
$\mathcal{CN}(0,\sigma^2)$ denotes a circularly symmetric complex Gaussian random variable with zero mean and variance  $\sigma^2$. 
$ \D_{\mathbf{a}} $ represents a diagonal matrix whose diagonal entries are elements from vector $\mathbf{a}$.

\section{System Model and Preliminaries} \label{sec:system_model}
We consider a \gls{dmimo} infrastructure with two \glspl{ap} A and B equipped with $M_{\tA}$ and $M_{\tB}$ antennas, respectively. 
We envision a scenario where both the \glspl{ap} do not have a strong communication link between them. 
There is a single-antenna \gls{afr} R that can exchange measurements between A and B. 
The system model is depicted in Fig.~\ref{fig:system_model}, where the signaling between the \glspl{ap} is blocked by a tall building. 
We assume a block-fading channel model and that the system operates in \gls{tdd} mode.  

All forward and reverse link channels between the nodes are reciprocal; however, with unknown reciprocity errors. 
Let $\t_\tA = [t_{\tA,1} ~ t_{\tA,2} ~ \dots ~ t_{\tA,M_\tA}]^\tT$ and $\rr_\tA = [r_{\tA,1} ~ r_{\tA,2} ~ \dots ~ r_{\tA,M_\tA}]^\tT$, where $t_{\tA,i}$ and $r_{\tA,i}$ are the transmit and receive complex gain vectors of the \gls{rf} chain of the $i$th antenna at \gls{ap}-A.
Similarly, we define $\{\t_\tB, \rr_\tB\}$ as the transmit and receive complex gain vectors of the \gls{rf} chains at \gls{ap} B, respectively, of appropriate-sized vectors. 
We assume that the \glspl{ap} A and B are individually reciprocity calibrated~\cite{vieira2021reciprocity,vieira2017reciprocity,shepard2012argos,ganesan2024beamsync}.
The transmit and receive gains of the repeater R are defined as $t_\tR$ and $r_\tR$, respectively. 
Note that, in practice, the repeater  could be realized through a dual-antenna repeater, calibrated explicitly for reciprocity.


\subsection{Reciprocity Calibration of a Single AP}\label{sec:reciprocity_calibration}
In this subsection, we briefly describe how individual reciprocity calibration 
within one of the APs, say AP-A, works.
Consider a single antenna \gls{ue} with $t_u$ and $r_u$ as transmit and receive complex gain values in its \gls{rf} chain.
Let $\h_{\tA} = [h_{\tA,1} ~ h_{\tA,2} ~ \dots h_{\tA,M_\tA}]^\tT $ be the reciprocal channel between \gls{ap}-A and \gls{ue}. 

To understand the concept of reciprocity, let us assume a noiseless scenario and without loss of generality, the \gls{ue} transmits a scalar value $1$ as an uplink pilot. 
In the uplink channel estimation phase, the $i$th antenna of \gls{ap}-A receives $r_{\tA,i} h_{\tA,i} t_u$. 
In the downlink, \gls{ap} conjugates the uplink received signal and transmits it.
At the \gls{ue}, the signal received from the $i$th antenna of \gls{ap} can be written as 
\begin{align}\label{eqn:signal_from_antenna_i}
    \nonumber
    y_i & = r_u h_{\tA,i} t_{\tA,i} (r_{\tA,i} h_{\tA,i} t_u)^* \\
        & = \lvert t_u \rvert \lvert r_u \rvert \lvert t_{\tA,i} \rvert \lvert r_{\tA,i} \rvert \lvert h_{\tA,i} \rvert^2  e^{(\angle t_{\tA,i} - \angle r_{\tA,i} )- (\angle t_u - \angle r_u)} . 
\end{align}
It is clear from \eqref{eqn:signal_from_antenna_i} that the signals transmitted by   different AP antennas have different phases when  received at  the \gls{ue}, which  causes the signals from different antennas to add up non-constructively. Reciprocity calibration of the individual \gls{ap} is necessary to enable \gls{cjt} from \gls{ap}. 

The authors in \cite{shepard2012argos,vieira2017reciprocity} developed a scheme for  individual reciprocity calibration of an \gls{ap} using bi-directional measurements between the antennas in the \gls{ap}. 
We briefly outline this  calibration process to provide some background for the rest of the paper. 
\Gls{ap}-A selects one of the antennas as the reference antenna. Let $t_{\tA}$ and $r_\tA$ be the complex gain in the transmit and receive \gls{rf} chain of this reference antenna. 
For every $i$th antenna, using a bi-directional measurement with the reference antenna, \gls{ap}-A obtains the reciprocity calibration coefficient $\frac{t_\tA}{r_\tA} \frac{r_{\tA,i}}{t_{\tA,i}}$. (Note that for the reference antenna, the reciprocity calibration coefficient is $1$.)
During the downlink transmission from antenna $i$, \gls{ap} multiplies the transmitted signal by this reciprocity calibration coefficient.
Thus, with reciprocity calibration, \eqref{eqn:signal_from_antenna_i} becomes 
\begin{align}\label{eqn:signal_from_antenna_i_after_calibration}
\nonumber
    y_i & = r_u h_{\tA,i} t_{\tA,i} \frac{t_\tA}{r_\tA}\frac{r_{\tA,i}}{t_{\tA,i}} (r_{\tA,i} h_{\tA,i} t_u)^* \\
        & = \frac{\lvert t_\tA \rvert}{\lvert r_\tA \rvert}\lvert t_u \rvert \lvert r_u \rvert \lvert r_{\tA,i} \rvert^2 \lvert h_{\tA,i} \rvert^2  e^{(\angle t_\tA - \angle r_\tA )- (\angle t_u - \angle r_u)} . 
\end{align}
From \eqref{eqn:signal_from_antenna_i_after_calibration}, the signals received from all the antennas of \gls{ap}-A at \gls{ue} have the same phase, and therefore  add up coherently.
Being in the same entity and driven by the same \gls{lo}, these reciprocity calibration coefficients remain relatively stable over time, thus requiring only infrequent recalibration.
In vector form, application of reciprocity calibration at \gls{ap}-A can be written as multiplying $\frac{t_\tA}{r_\tA} \D_{\rr_\tA} \D_{\t_\tA}^{-1} $ with the transmit signal, where $\D_{\rr_\tA}\triangleq\text{diag}(r_{\tA,1},\ldots,r_{\tA,M_\tA})$ and $\D_{\t_\tA}\triangleq\text{diag}(t_{\tA,1},\ldots,t_{\tA,M_\tA})$.

\subsection{CJT from two APs in the Absence of a Repeater}\label{sec:cjt_without_repeater}
In this subsection, we explain the requirement of phase synchronization between \glspl{ap} to enable coherent combination of signals at the \gls{ue}. 
We consider a simple noiseless scenario, where two \glspl{ap} A and B   communicate with a single-antenna \gls{ue}. 
We assume that both \glspl{ap} are \emph{individually} reciprocity calibrated as described in Sec.~\ref{sec:reciprocity_calibration}.
For \gls{ap}-B, we let $t_\tB$ and $r_\tB$ be the complex gains of the transmit and receive \gls{rf} chain of its reference antenna.
Let $\h_\tB  \in \CC^{M_\tB \times 1}$ be the reciprocal channel between \gls{ap}-B and the \gls{ue}. 

In the uplink, \glspl{ap} A and B obtain the following signal, through their receive \gls{rf} chain, from the \gls{ue}:
\begin{align}
    \y_\tA & = \D_{\rr_\tA} \h_\tA t_u \\
    \y_\tB & = \D_{\rr_\tB} \h_\tB t_u.
\end{align}
In the downlink, \glspl{ap} A and B conjugate their received uplink signals and transmit them to the \gls{ue} in an orthogonal manner. 
The signal received from \gls{ap}-A at the \gls{ue} is given by 
\begin{align}\label{eqn:signal_APA_UE}
    \nonumber
    y_\tA & = r_u \h_\tA^\tT \D_{\t_\tA} \left( \frac{t_\tA}{r_\tA} \D_{\rr_\tA} \D_{\t_\tA}^{-1} \right) \left( \D_{\rr_\tA} \h_\tA t_u\right)^* \\
    & = \frac{\Abs{t_\tA}}{\Abs{r_\tA}} \Abs{t_u} \Abs{r_u} \Norm{\h_\tA^\tT \D_{\rr_\tA}}^2 e^{(\angle t_\tA - \angle r_\tA )- (\angle t_u - \angle r_u)}.
\end{align}
(In practice, normalization is required to satisfy a power constraint, but we ignore that here for the sake of simplicity.)
Note that the term $\frac{t_{\tA}}{r_{\tA}} \D_{\rr_\tA} \D_{\t_\tA}^{-1}$ represents the reciprocity calibration applied by  the antennas at \gls{ap} A. 
Similarly, the signal received from \gls{ap} B at the \gls{ue} is given by
\begin{align}\label{eqn:signal_APB_UE}
    y_\tB = \frac{\Abs{t_\tB}}{\Abs{r_\tB}} \Abs{t_u} \Abs{r_u} \Norm{\h_\tB^\tT \D_{\rr_\tB}}^2 e^{(\angle t_\tB - \angle r_\tB )- (\angle t_u - \angle r_u)}.
\end{align}
Clearly, from \eqref{eqn:signal_APA_UE} and \eqref{eqn:signal_APB_UE}, the signals received from both \glspl{ap} at \gls{ue} have different phases and may not add constructively. 
Hence, it is necessary to phase-calibrate the two APs relative to one another. 

BeamSync is a phase calibration method proposed in~\cite{ganesan2024beamsync} to synchronize two \glspl{ap}. We briefly describe it below for the noiseless scenario.
Let $\H \in \CC^{M_\tA \times M_\tB}$ be the reciprocal channel between the \glspl{ap} A and B. 
The synchronization is done in multiple stages. In stage-I, \gls{ap} B beamforms a signal to A in a direction $\f\in\CC^{M_\tB\times 1}$; the signal received at A is given by
\begin{align}
    \nonumber
    \y_\tA & = \D_{\rr_\tA} \H \D_{\t_\tB} \left( \frac{t_\tB}{r_\tB} \D_{\rr_\tB} \D_{\t_\tB}^{-1}\right)  \f \\
    & = \frac{t_\tB}{r_\tB} \D_{\rr_\tA} \H \D_{\rr_\tB} \f
    . 
\end{align}
Let $\Tilde{\H} = \D_{\rr_\tA} \H \D_{\rr_\tB}$. 
In stage-II, A sends the conjugate of the signal received in stage-I, to B; the received signal is given by
\begin{align}
    \y_\tB = \frac{t_\tA}{r_\tA} \Tilde{\H}^\tT \left( \frac{t_\tB}{r_\tB} \Tilde{\H} \f \right)^*. 
\end{align}
Combining with the known beamforming vector $\f$, \gls{ap}-B forms the following statistic: 
\begin{align}
    \nonumber
    \f^\tT \y_\tB & = \frac{t_\tA}{r_\tA} \f^\tT \Tilde{\H}^\tT \left( \frac{t_\tB}{r_\tB} \Tilde{\H} \f \right)^* \\
    & = \frac{\Abs{t_\tA}}{\Abs{r_\tA}} \frac{\Abs{t_\tB}}{\Abs{r_\tB}} \Norm{\Tilde{\H}\f}^2 e^{(\angle t_\tA - \angle r_\tA )- (\angle t_\tB - \angle r_\tB)}.
\end{align}
By computing $\angle (\f^\tT \y_\tB) =(\angle t_\tA - \angle r_\tA )- (\angle t_\tB - \angle r_\tB)$, \gls{ap}-B can precompensate the phase difference between the \glspl{ap} A and B, and align the phase with A during the downlink transmission to \gls{ue}.

\section{Repeater Aided Over-the-Air Phase Synchronization}
\label{sec:cjt_with_repeater}

In Sec.~\ref{sec:cjt_without_repeater}, we assumed that bi-directional measurements between the \glspl{ap} are possible.
We now address the case when the channel between the \glspl{ap} is too weak to 
perform such bi-directional measurements, and show how
a repeater, R, can aid the synchronization process between the \glspl{ap}
(Fig.~\ref{fig:system_model}).

\subsection{CJT with Repeater for a Noiseless Case}
\label{sec:cjt_with_repeater_noiseless}
We first outline how the algorithm works in the special case of a noiseless system. 
Let $\g_\tA \in \CC^{M_\tA \times 1}$ and $\g_\tB \in \CC^{M_\tB \times 1}$ be the reciprocal channels, between A and R, and between B and R, respectively. 
To enable a \gls{cjt}, both A and B need to be phase-synchronized and hence,  A (or B) needs to estimate the phase of  $\frac{t_\tB r_\tA}{r_\tB t_\tA}$ (or $\frac{t_\tA r_\tB}{r_\tA t_\tB}$) and use it to calibrate during downlink transmission.

In the presence of a repeater R, the synchronization between the \glspl{ap} is carried out in multiple stages. First, a synchronization reference signal $x\in\CC$ is transmitted by   \gls{ap} A to R which receives 
\begin{align}\label{eqn:cjt_stage1}
    y_{\tR1}=r_\tR \g_\tA^{\tT} \D_{\t_\tA} \Brac{\frac{t_{\tA}}{r_{\tA}} \D_{\rr_\tA} \D_{\t_\tA}^{-1}}\f_\tA x. 
\end{align}
The terms in \eqref{eqn:cjt_stage1} are as follows:
$\mathbf{f}_\tA$ is the transmit beamformer applied by A; the term $\frac{t_{\tA}}{r_{\tA}} \D_{\rr_\tA} \D_{\t_\tA}^{-1}$ represents the reciprocity calibration applied by all antennas at \gls{ap} A; then the signal passes through the transmit \gls{rf} chain of \gls{ap}-A and is represented by $\D_{\t_\tA}$; then the signal is propagated through the reciprocal channel $\g_\tA$; and is received at R through the receive \gls{rf} chain of R whose receive gain is represented by $r_\tR$. 

Since $x$ is a scalar pilot symbol known to both the \glspl{ap}, we can set it to $1$ without loss of generality. 
Therefore, 
\begin{align}\label{eqn:toy_example_stage1}
    y_{\tR1}= \frac{t_{\tA}}{r_{\tA}} r_\tR \g_\tA^{\tT} \D_{\rr_\tA} \f_\tA.
\end{align}	
Upon receiving $y_{\tR1}$, R forwards it to \gls{ap} B which applies a beamformer $\f_\tB$ to obtain the complex scalar
\begin{align}\label{eqn:toy_example_stage2}
    \nonumber
    y_{\tB2} &= \f_\tB^{\tT} \D_{\rr_\tB} \g_\tB  \Brac{t_\tR y_{\tR1}} \\ 
    & = t_\tR r_\tR \frac{t_{\tA}}{r_{\tA}} \f_\tB^{\tT} \D_{\rr_\tB} \g_\tB \g_\tA^{\tT} \D_{\rr_\tA} \f_\tA.
\end{align}	

Next, \gls{ap}-B processes $y_{\tB2}$ to obtain a second reference signal $x_\tB$ by conjugating~\eqref{eqn:toy_example_stage2} as 
\begin{align}\label{eqn:ref_signal_with_conjugate}
    x_\tB = y_{\tB2}^{*}. 
\end{align}
\Gls{ap}-B transmits the reference signal $x_\tB$ with a beamformer $\mathbf{f}_\tB$ to R (by following a similar procedure as explained in \eqref{eqn:cjt_stage1}), which receives
\begin{align}\label{eqn:toy_example_stage3}
    y_{\tR3} = \frac{t_{\tB}}{r_{\tB}} r_\tR \g_\tB^{\tT} \D_{\rr_\tB} \f_\tB x_\tB. 
\end{align}
On receiving the signal~\eqref{eqn:toy_example_stage3}, R forwards it to A, which in turn applies a beamformer $\f_\tA$ to obtain the complex scalar 
\begin{align}\label{eqn:toy_example_stage4}
    \nonumber
    y_{\tA4} & = \f_\tA^{\tT} \D_{\rr_\tA} \g_\tA  \Brac{t_\tR y_{\tR3}} \\  \nonumber
    & =  \frac{t_{\tB}}{r_{\tB}} \Brac{\frac{t_{\tA}}{r_{\tA}}}^{*}\lvert \f_\tA^{\tT} \D_{\rr_\tA} \g_\tA \rvert^2 \lvert \f_\tB^{\tT} \D_{\rr_\tB} \g_\tB \rvert^2 \lvert t_\tR \rvert^2 \lvert r_\tR \rvert^2  \\
    & = c \cdot e^{(\angle t_\tB - \angle r_\tB )- (\angle t_\tA - \angle r_\tA) }, 
\end{align}
where 
\begin{align}
    c = \frac{\lvert t_\tA \rvert}{\lvert r_\tA \rvert} \frac{\lvert t_\tB \rvert}{\lvert r_\tB \rvert} \lvert \f_\tA^{\tT} \D_{\rr_\tA} \g_\tA \rvert^2 \lvert \f_\tB^{\tT} \D_{\rr_\tB} \g_\tB \rvert^2 \lvert t_\tR \rvert^2 \lvert r_\tR \rvert^2,
\end{align}
is a real scalar. 
By estimating the phase of the measurement in~\eqref{eqn:toy_example_stage4} and using the estimated quantity to correct the phase during the transmission by \gls{ap} A, both   \glspl{ap} can perform \gls{cjt} to the \gls{ue}.

This  demonstrates that phase synchronization is achievable in a noiseless scenario using bi-directional measurements routed through a repeater. 
In the following sections, we extend this over-the-air synchronization approach to a noisy environment, leveraging the repeater.

\subsection{Repeater-Aided CJT: Two-Stage Synchronization Procedure}
\label{sec:cjt_with_repeater_noisy}



With our repeater-aided two-stage \gls{ota} procedure, a primary \gls{ap} and a secondary \gls{ap} exchange bidirectional measurements between them via the repeater. We designate \gls{ap}-A as the primary AP, without loss of generality.

\subsubsection{Stage-I}
In the first stage, \gls{ap}-A transmits a synchronization signal $\x\in\CC^{L\times1}$ satisfying $~\Norm{\x}^2 = L$, 
to R, which amplifies it and repeats towards \gls{ap}-B. The received signals received at R and \gls{ap}-B are given by 
\begin{align}
	\y_{\tR1}^\tT & = \sqrt{\rhoA}\frac{t_\tA}{r_\tA} r_\tR \g_\tA^{\tT} \D_{\rr_\tA} \f_\tA \x^{\tT} + \w_{\tR1}  \\
	\y_\tB^{\tT} & = \sqrt{\rhoR} t_\tR \f_\tB^{\tT} \D_{\rr_\tB} \g_\tB \y_{\tR1}^\tT + \w_\tB^{\tT}, 
\end{align}
respectively, where $\f_\tA\in\CC^{M_\tA\times 1}$ and $\f_\tB\in\CC^{M_\tB \times 1}$ are the unit-norm transmit beamforming vector of \gls{ap}-A and receive combining vector at \gls{ap}-B, respectively, $\rho_\tA$ and $\rho_\tR$ are the transmit powers of \gls{ap}-A and R, respectively, and $\w_{\tR1}$ and $\w_\tB$ are the \gls{awgn} at R and \gls{ap}-B, respectively, whose entries are \gls{iid} as $\CN(0,\sigma^2)$. 

Let $\Tilde{\g}_\tA = \D_{\rr_\tA} \g_\tA$ and $\Tilde{\g}_\tB = \D_{\rr_\tB} \g_\tB$, be the effective channel vectors from \glspl{ap} A and B to the repeater R, respectively. 
As in \cite{ganesan2024beamsync}, the beamforming vectors $\f_\tA$ and $\f_\tB$ are chosen as the dominant directions of the effective channels  $\Tilde{\g}_\tA$ and $\Tilde{\g}_\tB$, respectively.
These beamforming vectors can be obtained by sending an omnidirectional pilot signal from the repeater R, and both  \glspl{ap} can estimate the dominant direction of signal reception by computing the \gls{svd} of the incoming signal, and taking the dominant
singular vector.

\subsubsection{Stage-II}
In the second stage, the \gls{ap}-B transmits the conjugate of the signal received during stage-I using the beamformer $\f_\tB$ (i.e., $\f_\tB\y_\tB^{\tH}$) to R, which then forwards it to \gls{ap}-A. The received signals at R and \gls{ap}-A are then given by 
\begin{align}
	\y_{\tR2}^\tT  & = \sqrt{\frac{L}{C}} \sqrt{\rhoB} \frac{t_\tB}{r_\tB} r_\tR\Tilde{\g}_\tB^{\tT} \f_\tB \y_\tB^{\tH} + \w_{\tR2}^\tT , \label{eqn:YR2}\\
    \Y_{\tA2}  & = \sqrt{\rhoR} t_\tR \Tilde{\g}_\tA \y_{\tR2}^\tT  + \W_{\tA2}^{\tT} ,\label{eqn:YA2}
\end{align}
respectively, where $\rho_\tB$ is the transmit power of \gls{ap}-B; $\w_{\tR2}$ and $\W_{\tA2}$ are \gls{awgn} at R and \gls{ap}-A whose entries are \gls{iid} $\CN(0,\sigma^2)$. The secondary \gls{ap}-B scales down its transmit signal by $C \triangleq \EE \left\{ \lVert \y_\tB \rVert^2\right\}$ to satisfy its average transmit power constraint.

Next,  \gls{ap}-A combines its received signal using a beamformer $\f_\tA$ and right multiplies with the synchronization signal $\x$ to obtain the test statistic 
\begin{align}\label{eqn:sync_test_statistic}
    \nonumber
    y & = \f_{\tA}^{\tT} \Y_{\tA2} \x \\
    & = c e^{(\angle t_\tB - \angle r_\tB )- (\angle t_\tA - \angle r_\tA)} + w,
\end{align}
where 
\begin{align}
    c & = \frac{L}{\sqrt{C}} \rhoR \sqrt{\rhoA\rhoB} \lvert t_\tR \rvert^2 \lvert r_\tR \rvert^2 \frac{\lvert t_\tA \rvert \lvert t_\tB \rvert}{\lvert r_\tA \rvert \lvert r_\tB \rvert } \left\lvert \f_\tA^{\tT} \Tilde{\g}_\tA \Tilde{\g}_\tB^{\tT} \f_\tB  \right\rvert^2 \label{eqn:constant_c}
\end{align}
is a positive scalar, and $w$ consists of all   composite \gls{awgn} terms. 

The \gls{ap}-A needs to estimate $\theta \triangleq (\angle t_\tB - \angle r_\tB )- (\angle t_\tA - \angle r_\tA)$ to enable \gls{cjt} along with \gls{ap}-B.  
The quantity $c$ being a real-valued scalar, 
an estimate of $\theta$ is to compute the phase of $y$, i.e.,
\begin{align}
    \hat{\theta} = \angle y. 
\end{align}
Note that, to estimate $\theta$, \gls{ap}-A does not require knowledge of the effective channels $\Tilde{\g}_\tA$ and $\Tilde{\g}_\tB$, nor does it need to know the reciprocity coefficients.
For a reliable estimate of $\theta$, we need the quantity $c$ to be as large as possible.

\subsection{Scalability}
Scalability in the context of repeater-aided synchronization could be achieved through the formation of synchronization clusters, as proposed in the partially coherent \gls{dmimo} framework~\cite{ganesan2024cell}. 
In particular, repeaters and  \glspl{ap} within their coverage could be incorporated into a common synchronization cluster, without increasing system complexity. 
Moreover, nodes that are initially unsynchronized could be progressively integrated, allowing sequential synchronization over time while preserving scalability.

\section{Results} \label{sec:results}
In this section, we present performance results for our proposed algorithm. 
We consider that the \glspl{ap} A and B are $d$ meters away from the repeater R, and that the \glspl{ap} do not have a direct link between them. Each element of the channels between each entity is modeled as an independent Rayleigh fading random variable with large-scale fading path-loss given by
\begin{align}
    \text{PL (in dB)} = -30.5-36.7\log_{10}(d). 
\end{align}
To compute the noise variance $\sigma^2$, we assume that the system is operating at a temperature of $290$~K, with a bandwidth of $20$~MHz and a noise figure of $9$~dB. 

\begin{figure}[!t]
    \centering
%
%
\begin{tikzpicture}[font=\scriptsize]

\begin{axis}[%
width=0.4\textwidth,
height=0.3\textwidth,
at={(0.758in,0.481in)},
scale only axis,
xmin=0,
xmax=80,
xlabel={$d$ meters},
ymode=log,
ymin=1e-06,
ymax=1,
yminorticks=true,
ylabel={RMSE (in radians)},
axis background/.style={fill=white},
xmajorgrids,
ymajorgrids,
yminorgrids,
legend style={at={(0.6,0.05)}, anchor=south west, legend cell align=left, align=left, draw=white!15!black}
]
\addplot [color=blue, line width=1.0pt, mark=o, mark options={solid, blue}]
  table[row sep=crcr]{%
1	6.67387634195056e-06\\
5	0.00013198209508648\\
10	0.000463431848464439\\
15	0.000980673463945883\\
20	0.00177480173000293\\
25	0.0026828490798988\\
30	0.00392895625896283\\
35	0.00560502360139451\\
40	0.00816985136353637\\
45	0.0109898027039776\\
50	0.0166378627806195\\
55	0.0233052311373355\\
60	0.0332498615422599\\
65	0.0491874752558027\\
70	0.11890272535305\\
75	0.253921853682063\\
80	0.326757417287834\\
85	0.470310335131916\\
90	0.682160048952359\\
95	0.803564444567194\\
100	0.964034854371387\\
};
\addlegendentry{$\rhoR$=1mW}

\addplot [color=red, dashed, line width=1.0pt, mark=diamond, mark options={solid, red}]
  table[row sep=crcr]{%
1	7.0350107553558e-06\\
5	0.000131645296852045\\
10	0.000471677344210151\\
15	0.000998115525054451\\
20	0.00174829622068029\\
25	0.00261825455680478\\
30	0.00384473268828374\\
35	0.00514685433741626\\
40	0.00675949374187185\\
45	0.00940153198328637\\
50	0.0124207221547173\\
55	0.0161925208025459\\
60	0.0210068338363762\\
65	0.0286519453304978\\
70	0.0411875261073899\\
75	0.0609531500904528\\
80	0.128249560238736\\
85	0.14512656908377\\
90	0.270323190293835\\
95	0.444418544808036\\
100	0.574107244823907\\
};
\addlegendentry{$\rhoR$=2mW}

\addplot [color=black, line width=1.0pt, mark=x, mark options={solid, black}]
  table[row sep=crcr]{%
1	6.6452721869364e-06\\
5	0.000130964376648316\\
10	0.000461204019653311\\
15	0.000962545848954975\\
20	0.00173084811416353\\
25	0.00254663708871275\\
30	0.00378881184852108\\
35	0.00489728362587536\\
40	0.0065530226389784\\
45	0.00881536935385356\\
50	0.0103619198705527\\
55	0.0127879116042\\
60	0.0166806993241237\\
65	0.0197347383847832\\
70	0.0245515069962638\\
75	0.0329705885358544\\
80	0.0401818076668703\\
85	0.0554324632433655\\
90	0.0629249328319724\\
95	0.10238234126913\\
100	0.13777170120011\\
};
\addlegendentry{$\rhoR$=5mW}

\addplot [color=blue, dashed, line width=1.0pt, mark=star, mark options={solid, blue}]
  table[row sep=crcr]{%
1	6.98175505153707e-06\\
5	0.000133743720160461\\
10	0.000475323752781384\\
15	0.00102240849977498\\
20	0.00164449120127599\\
25	0.00260778615124611\\
30	0.00359150493279504\\
35	0.00497943343557551\\
40	0.00592689047226298\\
45	0.00776214731690802\\
50	0.00984131285799058\\
55	0.0113535740794031\\
60	0.0147824919762989\\
65	0.0169541046335622\\
70	0.0197169931973991\\
75	0.0258242862664236\\
80	0.028908536140862\\
85	0.0356454389458184\\
90	0.042761242900264\\
95	0.0825103931184488\\
100	0.0625961365017507\\
};
\addlegendentry{$\rhoR$=10mW}

\end{axis}

\end{tikzpicture}%
    \caption{Performance of proposed algorithm for varying powers at the repeater. The simulation parameters are $M_\tA = M_\tB = 16$, $\rhoA = \rhoB = 100$mW, $L=10$.}
    \label{fig:rmse_distance_with_different_repeater_power}
\end{figure}
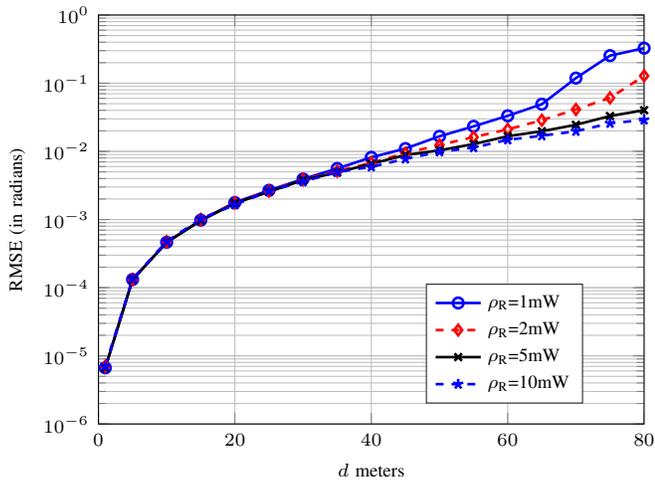

Fig.~\ref{fig:rmse_distance_with_different_repeater_power} shows the performance of the proposed algorithm as function of the distance $d$. 
For a $3$~GHz carrier signal, a \gls{rmse} of $0.01$~radians corresponds to an error of $0.5$~ps. 
This shows how we can correct for the carrier phase offset difference between the distributed \glspl{ap}. 
As the distance increases, the performance degradation could be mitigated by increasing the output power of the repeater.

\section{Conclusions} \label{sec:conclusions}

This paper presents a novel over-the-air algorithm designed for precise phase synchronization of distributed \glspl{ap} in \gls{dmimo} systems, addressing scenarios where direct communication links between the APs are not feasible. 
We develop a simple phase estimator that does not require \gls{csi} between any entities, nor does it require prior knowledge of any of the reciprocity coefficients. 
Numerical evaluation demonstrates the potential of the proposed algorithm, showcasing its promising performance in achieving accurate phase synchronization.

\bibliographystyle{IEEEtran}
\bibliography{references}

@STRING{IEEE_J_WCOM       = "{IEEE} Trans. Wireless Commun."}

@STRING{IEEE_M_COMM         = "{IEEE} Comm. Mag."}

@STRING{IEEE_J_SP         = "{IEEE} Trans. Signal Process."}

@STRING{IEEE_J_WCOML         = "{IEEE} Wireless Comm. Letters"}

@string{EURASIP = "{EURASIP} J. Wireless Commun. Net."}

@string{pimrc = "Proc. Int. Sym. Personal Indoor Mobile Radio Commun. (PIMRC)"}

@article{ganesan2024cell,
  title={Cell-free massive {MIMO} with multi-antenna users and phase misalignments: {A} novel partially coherent transmission framework},
  author={Ganesan, Unnikrishnan Kunnath and Vu, Tung Thanh and Larsson, Erik G},
  journal={IEEE Open J. Commun. Soc.},
  volume={5},
  pages={1639--1655},
  year={2024},
  month={Mar.},
  publisher={IEEE}
}

@article{carvalho2025network,
  title={Network-controlled repeater-{A}n introduction},
  author={Carvalho, Fco Italo G and Paiva, Raul Victor de O and Maciel, Tarcisio F and Monteiro, Victor F and Lima, Fco Rafael M and Moreira, Darlan C and Sousa, Diego A and Makki, Behrooz and {\AA}str{\"o}m, Magnus and Bao, Lei},
  journal={IEEE Commun. Standards Mag.},
  month={Aug.},
  year={2025},
  publisher={IEEE}
}

@INPROCEEDINGS{kaltenberger2010relative,
  author={Kaltenberger, Florian and Jiang, Haiyong and Guillaud, Maxime and Knopp, Raymond},
  booktitle={Proc. Future Network \& Mobile Summit}, 
  title={Relative channel reciprocity calibration in {MIMO/TDD} systems}, 
  year={2010},
  month={Mar.},
  address={Florence, Italy},    
  pages={1-10},
}

@inproceedings{ma2015channel,
  title={Channel estimation error and beamforming performance in repeater-enhanced massive {MIMO} systems},
  author={Ma, Yiming and Zhu, Dengkui and Li, Boyu and Liang, Ping},
  booktitle=pimrc,
  pages={672--677},
  month={Aug.},
  address={Hong Kong, China},
  year={2015}
}

@article{sara2025repeater,
    title={Achieving Distributed {MIMO} Performance with Repeater-Assisted Cellular Massive {MIMO}}, 
    author={Sara Willhammar and Hiroki Iimori and Joao Vieira and Lars Sundström and Fredrik Tufvesson and Erik G. Larsson},
    journal=IEEE_M_COMM,
    month={Feb.},
    year={2025},
    pages={1--13}, 
    publisher={IEEE}
}

@article{interdonato2019ubiquitous,
	title={Ubiquitous cell-free massive {MIMO} communications},
	author={Interdonato, Giovanni and Bj{\"o}rnson, Emil and Quoc Ngo, Hien and Frenger, P{\aa}l and Larsson, Erik G},
	journal=EURASIP,
	volume={2019},
	number={197},
	pages={1--13},
	month={Dec.},
	year={2019},
	publisher={Springer}
}

@article{ngo2017cell,
	title={Cell-free massive {MIMO} versus small cells},
	author={Ngo, Hien Quoc and Ashikhmin, Alexei and Yang, Hong and Larsson, Erik G and Marzetta, Thomas L},
	journal=IEEE_J_WCOM,
	volume={16},
	number={3},
	pages={1834--1850},
	year={2017},
	month={Mar.},
	publisher={IEEE}
}

@article{rogalin2014scalable,
	title={Scalable synchronization and reciprocity calibration for distributed multiuser {MIMO}},
	author={Rogalin, Ryan and Bursalioglu, Ozgun Y and Papadopoulos, Haralabos and Caire, Giuseppe and Molisch, Andreas F and Michaloliakos, Antonios and Balan, Vlad and Psounis, Konstantinos},
	journal=IEEE_J_WCOM,
	volume={13},
	number={4},
	pages={1815--1831},
	year={2014},
	month={Apr.},
	publisher={IEEE}
}

@article{vieira2017reciprocity,
	title={Reciprocity calibration for massive {MIMO}: {P}roposal, modeling, and validation},
	author={Vieira, Joao and Rusek, Fredrik and Edfors, Ove and Malkowsky, Steffen and Liu, Liang and Tufvesson, Fredrik},
	journal=IEEE_J_WCOM,
	volume={16},
	number={5},
	pages={3042--3056},
	month={May},
	year={2017},
	publisher={IEEE}
}

@inproceedings{vieira2021reciprocity,
	title={Reciprocity calibration of Distributed Massive {MIMO} Access Points for Coherent Operation},
	author={Vieira, Joao and Larsson, Erik G},
	booktitle=pimrc,
	pages={783--787},
    address={Helsinki, Finland},
	month={Sep.},
	year={2021}
}

@article{nissel2022correctly,
	title={Correctly Modeling {TX} and {RX} Chain in (Distributed) Massive {MIMO}-New Fundamental Insights on Coherency},
	author={Nissel, Ronald},
	journal=IEEE_J_WCOML,
	year={2022},
	month={Oct.},
	volume={26},
	number={10},
	pages={2465-2469},
	doi={10.1109/LCOMM.2022.3189968},
	publisher={IEEE}
}

@inproceedings{shepard2012argos,
	title={Argos: {P}ractical many-antenna base stations},
	author={Shepard, Clayton and Yu, Hang and Anand, Narendra and Li, Erran and Marzetta, Thomas and Yang, Richard and Zhong, Lin},
	booktitle={Proc. Int. Conf. Mobile Computing Networking (MobiCOM)},
	pages={53--64},
	month={Aug.},
    address={ Istanbul, Turkey},
	year={2012}
}

@article{ganesan2024beamsync,
	author={Ganesan, Unnikrishnan Kunnath and Sarvendranath, Rimalapudi and Larsson, Erik G.},
	journal=IEEE_J_WCOM, 
	title={BeamSync: {O}ver-The-Air Synchronization for Distributed Massive {MIMO} Systems}, 
	year={2024},
        month={Jul.},
        volume={23},
	number={7},
	pages={6824--6837},
	doi={10.1109/TWC.2023.3335089}
}

@article{larsson2024massive,
	author={Larsson, Erik G.},
	journal=IEEE_J_SP, 
	title={Massive Synchrony in Distributed Antenna Systems}, 
	month={Jan.},
	year={2024},
	volume={72},
	number={},
	pages={855-866},
	doi={10.1109/TSP.2024.3358618}
}

@ARTICLE{larsson2023phase,
	author={Larsson, Erik G. and Vieira, Joao},
	journal=IEEE_J_WCOML, 
	title={Phase Calibration of Distributed Antenna Arrays}, 
	month={Jun.},
	year={2023},
	volume={27},
	number={6},
	pages={1619-1623},
	doi={10.1109/LCOMM.2023.3266836}
}

\end{document}